\documentclass[twocolumn,english]{revtex4-1}
\usepackage[T1]{fontenc}
\usepackage[latin9]{inputenc}
\usepackage{color}
\usepackage{appendix}
\usepackage{babel}
\usepackage{longtable}
\usepackage{amsmath}
\usepackage{graphicx}
\usepackage[normalem]{ulem}
\usepackage[unicode=true,pdfusetitle,
 bookmarks=true,bookmarksnumbered=false,bookmarksopen=false,
 breaklinks=false,pdfborder={0 0 1},backref=false,colorlinks=true]
 {hyperref}
\begin{document}
\title{Particle-in-Cell Simulation of the Parametric Decay Instability of Alfv\'en Waves with Absorbing Boundary Conditions}
\author{Vijay Shankar$^1$, Feiyu Li$^1$, Seth Dorfman$^{2,3}$, and Xiangrong Fu$^{1,4}$}
\affiliation{$^1$New Mexico Consortium, Los Alamos, NM 87544 USA}
\affiliation{$^2$Space Science Institute, Boulder, Colorado 80301, USA}
\affiliation{$^3$University of California, Los Angeles, Los Angeles, California 90095}
\affiliation{$^4$Los Alamos National Laboratory, Los Alamos, New Mexico 87545, USA}
\begin{abstract}
The Alfv\'en wave parametric decay instability (PDI) facilitates energy transfer, plasma heating, and turbulence generation in space, astrophysical, and fusion plasmas. Most simulation studies of Alfv\'en wave PDI have focused on kinetic ions under periodic boundary conditions. Here, we present fully kinetic one-dimensional simulations (perpendicular wave-vector $k_\perp=0$) of the Alfv\'en wave PDI at low plasma beta using absorbing boundary conditions for the waves to understand the energy partition in an open system. For $\beta=5\times 10^{-4}$ and a normalized wave amplitude $\frac{\delta B}{B_0}=0.01$, nearly 92\% of the pump wave energy is transferred to the backward-propagating child Alfv\'en wave, and the remaining energy is partitioned between electrons ($\sim 1$-$2\%$) and ions ($\sim 6$-$7\%$). 
% Clear electron heating due to PDI is observed; however, the fraction of pump energy converted into electron thermal energy is small, $\sim 1$-$2\%$. 
% We observe that, 
In the parameter regime considered,
% \textcolor{blue}{(typical of the Large Plasma Device at UCLA)}, 
the ion and electron heating appears only when the PDI has sufficiently developed, and their rates are approximately twice the linear PDI growth rate, which roughly corresponds to the quadratic dependence of energy on the fluctuation amplitude. 
% , and suggests that PDI plays a significant role in converting wave energy into particle heating.
Furthermore, we find a qualitative agreement between theoretical and numerical growth rates over a range of plasma and wave parameters. 
This work establishes critical steps for future extension to finite $k_\perp$ waves in high dimensions, where stronger electron heating may be induced.  
\end{abstract}
\maketitle
\section{Introduction} 
Alfv\'en waves (AWs) are fundamental plasma modes found in space \cite{belcher_large-amplitude_1971,knudsen_alfven_1992,de_pontieu_chromospheric_2007,kasper_alfvenic_2019,lysak_kinetic_2023}, astrophysical \cite{hollweg_kinetic_1999,gonzalez_particle--cell_2023}, and laboratory plasmas \cite{leneman_laboratory_1999,wilcox_experimental_1960,gekelman_review_1999,chen_physics_2016,gekelman_many_2011,carter_laboratory_2006,howes_toward_2012}. They are believed to play a key role in a wide range of phenomena, including solar coronal heating \cite{1983,shoda_frequency-dependent_2018}, solar wind acceleration \cite{belcher_large-amplitude_1971,erdelyi_are_2007,verdini_turbulence_2009}, minor ion heating in solar flares \cite{fu_heating_2020}, redistribution and transport of energy in various plasmas conditions \cite{qiu_nonlinear_2018,qiu_spontaneous_2013,chen_physics_2016}. AWs can become unstable to various parametric instabilities \cite{Taylor_1970,li_hybrid_2022,ishizaki_parametric_2024,dorfman_measurement_2026,hasegawa_kinetic_1976,wong_parametric_1986}, among which the parametric decay instability (PDI) is one of the most common,  leading to the decay of a large-amplitude forward-propagating pump AW into a backward-propagating AW and a forward-propagating ion-acoustic wave. The ion-acoustic mode is driven by parallel ponderomotive forces \cite{chen1984introduction} arising from the nonlinear coupling of the two AWs. In fusion plasmas, toroidal Alfv\'en eigenmodes (TAEs) may nonlinearly decay into lower-frequency TAEs and geodesic acoustic modes, thereby influencing turbulence and transport in the fusion reactors \cite{qiu_nonlinear_2018,qiu_spontaneous_2013,zhu_nonlinear_2022,shankar_influence_2025}. These studies highlight the broad relevance of AW PDI across plasma environments.

Beyond its general importance, PDI exhibits similar behavior in both space and laboratory plasmas. In particular, plasma conditions in solar coronal holes and in the Large Plasma Device (LAPD) at UCLA share comparable dimensionless parameters, suggesting a useful space-laboratory correspondence. A representative set of dimensionless parameters-including plasma beta ($\beta$), normalized wave amplitude, and perpendicular scale parameters ($k_\perp^2\rho_s^2$ and $k_\perp^2\rho_i^2$)-is summarized in Table I of Ref.~\cite{bose_measured_2019}. Here, $\beta$ denotes the ratio of thermal to magnetic pressure, while $\rho_s$ and $\rho_i$ represent the ion-sound and ion gyroradii, respectively. Under these dimensionless parameter conditions, similar wave dispersion relations arise in both solar coronal hole and LAPD plasmas. In both cases, $k_\perp^2\rho_s^2\ll1$ and $k_\perp^2\rho_i^2\ll1$, although the ratio $k_\perp/k_\parallel$ can remain finite \cite{morales_structure_1994,morales_structure_1997,zhao_kinetic_2011,kiyani_dissipation_2015}, where $k_\parallel$ is the parallel wavenumber. Therefore, the dependence of the parametric decay instability (PDI) on perpendicular scales is an important aspect to investigate.

\indent The PDI of Alfv\'en waves has been extensively studied through theoretical \cite{derby1978modulational,goldstein_instability_1978,Taylor_1970} and numerical investigations  \cite{li_effects_2024,li_hybrid_2022,gonzalez_particle--cell_2023,fu_parametric_2018,gonzalez_proton_2021,gonzalez_role_2020,li_measuring_2025,li_parametric_2022,ishizaki_parametric_2024}. Laboratory experiments have also explored this process; however, direct observation of PDI driven by a single pump wave remains challenging. In LAPD, the addition of a second counter-propagating Alfv\'en wave enabled the nonlinear excitation of ion-acoustic modes \cite{dorfman_nonlinear_2013}; the growth rate of the parametric decay instability (PDI) was subsequently measured by observing the reduced spatial damping of a small counter-propagating (seed) Alfv\'en wave in the presence of a large pump wave \cite{dorfman_measurement_2026}. 

The PDI results in the transfer of pump-wave energy to both waves and particles \cite{gonzalez_particle--cell_2023}. Despite extensive work, the kinetic mechanisms governing energy transfer and dissipation in PDI remain poorly understood. In particular, the fraction of pump-wave energy converted into internal energy, as well as its partitioning between ions and electrons, remains an open question. This is central to understanding PDI-driven heating and acceleration in the solar wind, as well as heating in fusion plasmas.
Most previous simulations have primarily considered hybrid models, treating ions kinetically and electrons as a fluid, under both periodic \cite{gonzalez_proton_2021,gonzalez_role_2020} and open-boundary conditions \cite{li_effects_2024,li_hybrid_2022,li_parametric_2022,li_measuring_2025}. However, electrons can also gain energy during PDI through interactions with ion-acoustic waves and AWs \cite{gonzalez_particle--cell_2023}. This necessitates fully kinetic treatments in which both species are modeled as particles. Although a few such studies exist, they have largely been restricted to periodic domains \cite{gonzalez_particle--cell_2023,sakai_particle_2005,nariyuki_parametric_2008}.

Under periodic boundary conditions, waves repeatedly interact with a localized plasma, which can lead to an overestimation of energy partitioning and introduce additional artifacts, including discrepancies with finite, non-periodic wave evolution and the potential misrepresentation of instability signatures in comparisons with observations, as discussed in Ref. \cite{li_parametric_2022}. Furthermore, linear analysis presented in Ref. \cite{marriott_parametric_2024} shows that AWs are approximately $3-5$ times more stable under open boundary conditions than under periodic ones. This is attributed to the reduced growth rates in open systems by a similar factor, whereas periodic systems exhibit enhanced growth due to repeated wave-plasma interactions. By contrast, absorbing boundary conditions for waves are more realistic, as they prevent repeated wave-plasma interactions and enable a more accurate determination of energy transfer between ions and electrons. This motivates the present study of the PDI of AWs using fully kinetic simulations with absorbing boundary conditions for the waves. Moreover, establishing and understanding the 1D limit ($k_\perp = 0$) provides a necessary baseline before extending to higher-dimensional simulations. While the present work focuses on this 1D framework, it serves as a foundation for future studies with finite $k_\perp$, where stronger electron heating and its feedback on PDI are expected to play a significant role in more realistic 2D/3D configurations.

In this work, we perform fully kinetic simulations of the parametric decay instability of a left-hand circularly polarized (LHCP) Alfv\'en wave using the open-source SMILEI particle-in-cell code \cite{derouillat_smilei_2018}. The simulations are carried out in one spatial dimension with absorbing boundary conditions for the waves, while particles are confined by reflecting boundaries. We quantify the partitioning of pump-wave energy into backward-propagating waves and particle populations under plasma conditions relevant to the LAPD. The energy transfer to daughter modes and the resulting partitioning between electrons and ions are systematically examined over a range of wave and plasma parameters. To our knowledge, this is the first fully kinetic study of AW PDI with absorbing wave boundary conditions.\\

The remainder of the paper is organized as follows. The numerical simulation setup is described in Section \ref{sec:parameters}. The main results are presented in Section \ref{sec:results}, and a summary of the findings is given in Section \ref{sec:summary}. A few supplementary materials are also provided at the end of the paper in Section \ref{sec:sup_mat}.\\

\section{Numerical Scheme and Simulation Parameters \label{sec:parameters}}
The schematic of the simulation is shown in Fig.~\ref{fig:space}(a), where the domain is divided into three regions: (a) the field mask, indicated by the grey shaded area; (b) the pump AW injection region, shown in green; and (c) the physical region. The mask regions extend from $0-25 d_i$ \& $87-112 d_i$, and the wave injection region from $30-32d_i$. Two virtual walls are placed at $z_i=32 d_i$ and $z_f=87 d_i$ to restrict the particles' motion beyond these points, which is indicated by the vertical black dashed lines in Fig.~\ref{fig:elsseser}(a). All spatial scales in this manuscript are normalized to the ion inertial length $d_i=c/\omega_{pi}$ with $c$ and $\omega_{pi}$ representing the speed of light and ion plasma frequency\\ 

To implement absorbing boundary conditions for the waves, field masks are applied symmetrically at both ends of the simulation domain by modifying the transverse electric ($E$) and magnetic ($B$) fields in the code. The $z$ direction is taken to be parallel ($\parallel$) to the ambient magnetic field ($B_0$), while the $x$ and $y$ directions are perpendicular ($\perp$) to it. The masking scheme follows that used in Refs. \cite{umeda_improved_2001,li_parametric_2022}, and is given as follows:

\begin{equation}
F_m(z) =
\begin{cases}
1-\left(r\frac{z-d_z}{d_z}\right)^2, & z \le d_z \\
1,  & d_z \le z \le (L_z-d_z) \\
1-\left(r\frac{z-(L_z-d_z)}{d_z}\right)^2,   & (L_z-d_z) \le z \le L_z
\end{cases}
\end{equation}
Here, $L_z$, $d_z$, and $r$ denote the simulation domain length, field mask length, and the masking parameter that defines the gradient of the mask function ($F_m(z)$). It should be noted that the fields along the z-direction are not masked to avoid instabilities arising from background field gradients; additionally, the mean field is retained to ensure proper wave propagation along the z-direction.\\

An LHCP pump AW has been injected using the Antenna module of the code. The previous studies \cite{li_parametric_2022,gonzalez_particle--cell_2023,li_hybrid_2022,li_effects_2024} have prescribed $B$ alone or combinations of $B$ and $E$ or $B$, $E$, and $v$ fields for wave injection into the plasma. While such prescriptions are commonly used in simulations, antenna-based wave injection provides a more realistic description for laboratory plasmas, such as the Large Plasma Device (LAPD). Therefore, in this work, we prescribe the current density profiles for wave injection, which are given as follows,
\begin{align}
    \delta \vec{J}=\delta J_0 [ \sin(\omega_+ t) \hat{x}+ \cos(\omega_+ t)\hat{y}]\chi(t) \sigma(z).
\end{align}
Here, $\delta J_0$ and $\omega_+$ represent the current density amplitude and pump AW frequency, respectively. To induce a gradual injection of the pump AW into the plasma, the temporal envelope $\chi(t)$ ramps up as $\sin^2(\pi t/2t_{r})$ for $ t\leq t_r=50\Omega_{ci}^{-1}$ and then keeps constant, where $\Omega_{ci}=eB_0/m_i$ denotes the ion cyclotron frequency with $e$ being the elementary charge and $m_i$ the ion mass. The antenna current has a flat-top profile $\sigma(z)=1$ for $30d_i \le z \le32 d_{i}$ (and $\sigma(z)=0$ elsewhere). The length $a_w=2d_i$ is chosen to maximize the power transmission by the antenna; see Appendix ~\ref{sec:sup_mat} for details. 
% which is obtained by testing different values of $a_w$. \textcolor{blue}{Further details of choosing $a_w$ can be found in the .} 
The $E$ and $B$ fields evolve self-consistently from the antenna current according to Maxwell's equations.  We use $\omega_+=0.63 \Omega_{ci}$ and $\delta J_0=0.01/ev_An_0$, where $v_A$ and $n_0$ denote the Alfv\'en speed, and background plasma density, respectively. Our analysis focus on the central region of the simulation domain bounded by virtual walls at $z_i$ and $z_f$.
% , where the PDI develops. 
These virtual walls reflect particles back into the central region upon impact while allowing waves to pass through. The purpose of these walls is to prevent 
% particle loss from the central region and 
particle intervention from the mask regions, thereby enabling an accurate estimation of electron and ion heating resulting from PDI. \\

\indent The simulations are performed with speed ratio $c/v_A=\omega_{pi}/\Omega_{ci}=20$, ion and electron mass ratio $m_i/m_e=100$, total plasma beta $\beta=5\times 10^{-4}$, electron and ion temperature ratio $T_e/T_i=4.0$, spatial grid resolution $\Delta z=0.03125 d_i$, time step $\Delta t=0.030625/\omega_{pi}$, $L_z=112d_i$, $d_z=25d_i$, $r=0.04$, and $3000$ particles per cell. It should be noted that the reduced speed and mass ratio were adopted to save computation, and the chosen values are sufficient to capture the physical phenomena under investigation \cite{verscharen_dependence_2020}. The particles follow the Maxwellian distribution with reflective boundary conditions at both ends of the simulation domain. The ion plasma frequency $\omega_{pi}$ is used as the reference frequency. The initial ion and electron temperatures have been calculated using $T_e=[0.25\beta_e (m_i/m_e)/{(\omega_{pi}/\Omega_{ci})^2}] K_r$  and $T_i=[0.25\beta_i (m_i/m_e)/{(\omega_{pi}/\Omega_{ci})^2}] K_r$ while parallel magnetic field $B_0$ has been obtained using $B_0=(m_i/m_e)/(\omega_{pi}/\Omega_{ci}) B_r$ and the plasma density $n_0=(m_i/m_e) N_r$. Here, $K_r$, $B_r$, and $N_r$ denote the reference energy, magnetic field, and plasma density, respectively, as used in the SMILEI code, which are used to get the normalized form of Maxwell's equations. Collisionless plasma is assumed in this study. Ion and electron plasma beta values, $\beta_i$ and $\beta_e$, are calculated as $\beta_i=\beta/(1+T_e/T_i)$ and $\beta_e=\beta/(1+T_i/T_e)$, respectively. 

\section{Simulation Results \label{sec:results}}

\begin{figure}
\includegraphics[width=8cm]{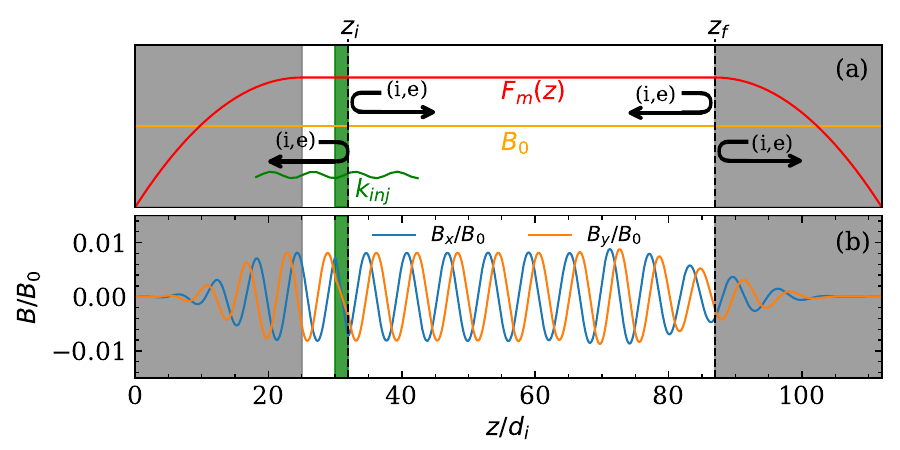}
\caption{\label{fig:space} Panel (a) represents the simulation schematic and panel (b) represents the transverse components of the magnetic fields at t=$200\Omega^{-1}_{ci}$. The grey-shaded region indicates the mask, while the green shows the wave injection. The two vertical black dashed lines denote the virtual walls used to restrict particle motion beyond the central region.}
\end{figure}

In this section, we first discuss the simulation results for a set of simulation parameters provided in Sec.\ref{sec:parameters}. We then follow with the parameter scans ($\delta B/B_0$, $\beta$, and $T_e/T_i$) around these base conditions. Fig.~\ref{fig:space}(b) shows the perpendicular magnetic fields ($B_x$ and $B_y$) at $t=200\Omega^{-1}_{ci}$. It is observed that both components are sufficiently damped within the mask region, with negligible wave reflection, indicating the proper implementation of the field mask. Before presenting the simulation results, we compare the theoretical and numerical linear dispersion relations. The theoretical dispersion relation of a LHCP AW with $k_\perp=0$ for a two-fluid plasma \cite{sakai_particle_2005, gonzalez_particle--cell_2023} in the low-frequency limit can be written as~\cite{hollweg_beat_1994}
\begin{align}
    \frac{\omega}{k}= v_A \sqrt{1-\frac{\omega}{\Omega_{ci}} }.\label{eq:disp}
\end{align}
The dispersion curve obtained from the simulation is found to be in good agreement with the linear dispersion relation given in Eq.~\ref{eq:disp}. The theoretical and numerical dispersion curves are presented in Fig.~\ref{fig:disp}(a) in the supplementary material. Although we have not prescribed the pump AW amplitude $\delta B/B_0$ for the initialization of the simulation, it may be estimated theoretically from the prescribed current density $\delta J_0 $ using Ampere's law $|\delta B|\sim\mu_0|\delta J_0|/ k_+$. Here, $k_+$ and $\mu_0$ represent the pump wavenumber and magnetic permeability, respectively. The wavenumber $k_+$ can be obtained from Eq.~\ref{eq:disp}; in particular, for $\omega_+=0.63\Omega_{ci}$, $k_+\sim 1 d_i^{-1}$. Consequently, $|\delta B|\sim 0.01B_0$.  \\

\begin{figure}
\includegraphics[width=8cm]{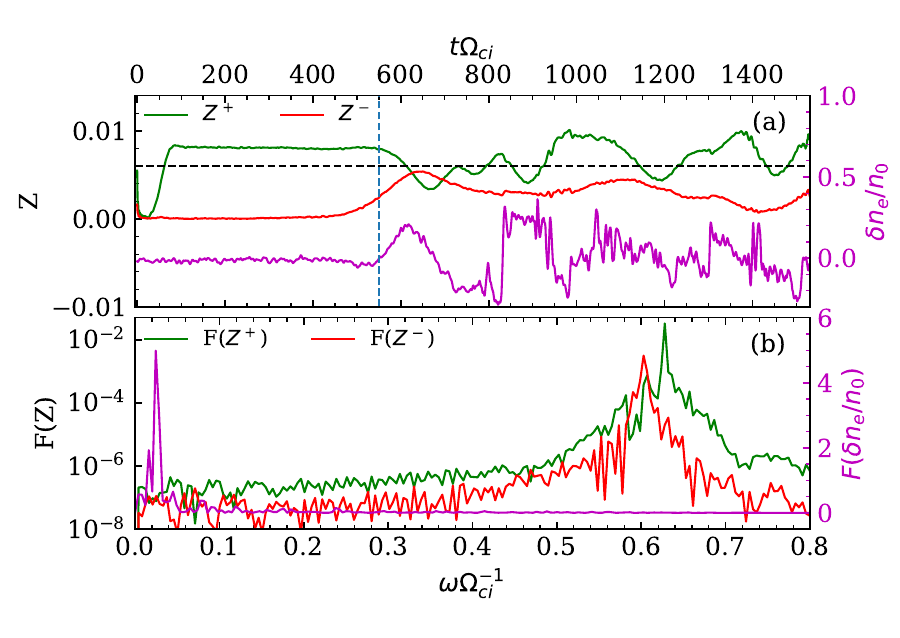}
\caption{\label{fig:elsseser}Panel (a) displays the forward-propagating pump AW ($Z^+$), the ion acoustic wave ($\delta n_e/n_0$), and the backward-propagating AW ($Z^-$) at a spatial location $z=40d_i$. Panel (b) presents the corresponding spectral densities with a time window of $1.02 \Omega_{ci}^{-1}$, where distinct peaks are observed for each of these modes. The vertical blue dashed line in panel (a) marks the onset time of PDI, while the black dashed horizontal line indicates the threshold pump amplitude required for PDI.} 
\end{figure}

\indent Due to PDI, the forward propagating pump AW decays into two daughter waves: a backward propagating AW and a forward propagating ion acoustic wave. The envelope of the forward ($Z^+$) and backward ($Z^-$) propagating AWs have been shown in Fig.~\ref{fig:elsseser} (a) in green and red colors, respectively, are obtained using the modified Els\"{a}sser's variables defined as $Z^\pm=(R_{bv}\delta v_y\mp {\delta B}/{\sqrt{\mu_0 \rho}})/{2v_A}$ \cite{li_hybrid_2022} at a spatial location $z=40d_i$; $\rho=m_in_0$, $\delta v_y$, and $R_{bv}$ denote the mass density, ion velocity field and ratio $(\delta B/B_0)/(\delta v_y/v_A)$, respectively. 
We have obtained $R_{bv}=0.61$ using $R_{bv}=(\delta B/B_0)/(\delta v_y/v_A)=(\omega_+/(k_+{v_A}))=\sqrt{1-\omega_+/\Omega_{ci}}$ \cite{hollweg_kinetic_1999}. The fluctuations in the electron density ($\delta n_e/n_0$) are also presented in Fig.~\ref{fig:elsseser}(a) by the magenta color curve. The ion density fluctuations exhibit characteristics similar to those of the electron density and are therefore not shown. The fluctuations in $Z^-$ and $\delta n_e/n_0$ appear around $t\Omega_{ci}=550$, indicating the onset of PDI. This time is marked by vertical dashed lines in Fig.~\ref{fig:elsseser}(a). The onset of the PDI requires a threshold pump amplitude given by $\delta B/B_0>4\beta^{3/4}\sqrt{T_i/T_e}$ \cite{li_hybrid_2022}. When $\delta B/B_0$ drops below the threshold  (black dashed horizontal line in Fig.~\ref{fig:elsseser}(a)), PDI is suppressed, and it reappears once the threshold is exceeded again. This behavior gives rise to multiple cycles in $Z^+$, $Z^-$, and $\delta n_e/n_0$ as illustrated in Fig.~\ref{fig:elsseser}(a). It is observed from Fig.~\ref{fig:elsseser}(a) that the amplitude of $Z^-$ is only slightly smaller than that of $Z^+$  {around $t\Omega_{ci}\sim 600-900$}, indicating that most of the pump-wave decays into the counter-propagating AW. The ion acoustic wave frequency, $\omega_s$, can be obtained by measuring the frequency difference between forward and backward AWs under the frequency-matching condition as $\omega_s=\omega_+-\omega_-\sim0.025\Omega_{ci}$. For this purpose, a Fourier transform of $Z^+$, $Z^-$, and $\delta n_e/n_0$ is performed, as shown in Fig.~\ref{fig:elsseser}(b). Here, $\omega_-$ denotes the backward AW frequency. Theoretically, $\omega_{sth}\sim2\sqrt{\beta} \omega_+=0.028\Omega_{ci}$ for the present set of parameters, which is similar to the simulation value. 

\begin{figure}
\includegraphics[width=8cm]{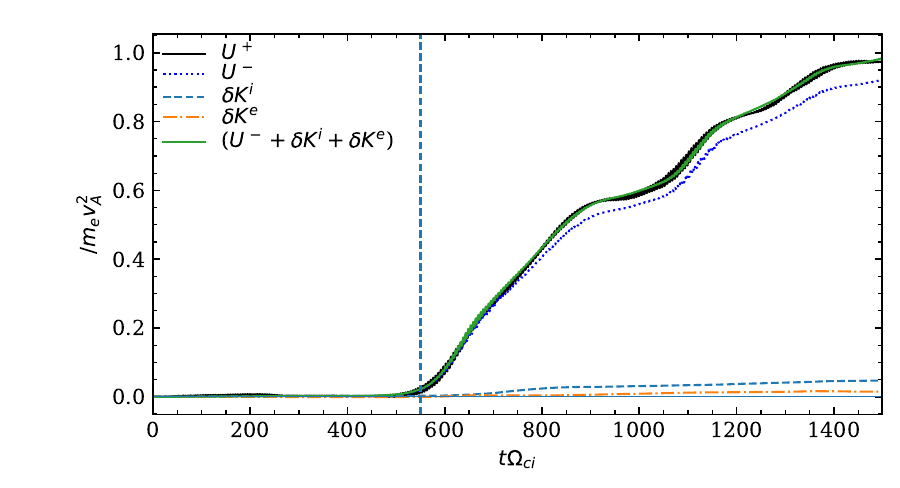}
\caption{\label{fig:energy_p} Energy partition of the pump AW among electrons, ions, and the counter-propagating AW. The sum of the ion, electron, and backward-wave energies (green curve) is nearly equal to the pump energy. The blue solid horizontal line indicates zero energy.}
\end{figure}

\indent Next, we analyze the energy partition of the pump wave into the counter-propagating AW wave and the ion and electron thermal energies. The energy  of the forward and backward AWs can be calculated by integrating $E^\pm=\frac{1}{2} |Z^\pm|^2$ over a certain time/spatial extent. The net energy loss of the pump wave ($U^+$) and the gain in the backward wave ($U^-$) energies are shown in Fig.~\ref{fig:energy_p} by black solid and blue dotted lines, respectively. The quantities $U^+$ and $U^-$ are computed using the following expression:
\begin{align}
      U^+(t) = {v_g^+}\int_{0}^{t} E^+(t')|_{z_i}dt' &-v_g^+\int_{0}^{t} E^+(t')|_{z_f}dt' \nonumber & \\  - \int_{z_i}^{z_f} E^+(z)|_{t}dz,\label{eq:uf}
\end{align}   

\begin{align}
     U^-(t) = {v_g^-}\int_{0}^{t} E^-(t')|_{z_i}dt' + \int_{z_i}^{z_f} E^-(z)|_{t}dz. \label{eq:ub}
\end{align}
Here, $t'$ runs over all previous times from $0$ to $t$. The variables $v_g^+$ and $v_g^-$ represent the group velocities of the forward and backward AWs, respectively. {The forward group velocity, $v_g^+$, has been calculated from the linear dispersion relation (Eq.~\ref{eq:disp}) as 
\begin{align}
    \frac {v_g^{\pm}}{v_A}=\frac{2(1-\omega_{\pm}/\Omega_{ci})^{3/2}}{(2-\omega_{\pm}/\Omega_{ci})}.
    \label{vgvg}
\end{align}
By contrast, $v_g^-$ has been calculated using the group velocity for backward AW derived from  Eq.~17 of Ref.\cite{derby1978modulational}. The estimate of $v_g^-$ from Eq.~\ref{vgvg} is slightly higher ($\sim 5\%$), as it does not self-consistently account for PDI contribution.} In Eq.~\ref{eq:uf}, the first and second terms represent the total pump energy entered into the central region through $=z_i$ and the total pump energy exited from the central region through $z=z_f$, while the third term gives the space-integrated pump energy from $z_i$ to $z_f$ at a given time $t$. Ideally, in the absence of a source or sink of energy in the central region, $U^+$ should be zero, and in the presence of any dissipation mechanism, it should be positive; therefore, $U+$ represents the net loss of the pump energy in the system due to PDI. Similarly, Eq.~\ref{eq:ub} denotes the gain in the backward wave energy ($U^-$). Figure~\ref{fig:energy_p} shows that $U^+$ and $U^-$ are zero initially, and they increase as PDI starts growing, indicating decay of pump-wave energy into the counter-propagating AW. Initially, all the pump energy is transferred to the backward-propagating child wave. However, as the acoustic wave becomes sufficiently developed {around $t\Omega_{ci}\sim750$}, {a clear energy gap between $U^+$ and $U^-$ begins to emerge}. {The difference of $U^+$ and $U^-$} is likely transferred to the electrons and ions. Therefore, we calculate the increase in ion and electron thermal energies $\delta K^i$ and $\delta K^e$ (with their respective initial value subtracted), represented by blue dashed and orange dot-dashed curves in Fig.~\ref{fig:energy_p}.  It is observed that  $\sim92$ \% of $U^+$ goes to $U^-$, $\sim6$\% of $U^+$ goes to $\delta K^i$, and  $\sim1.5$\% of $U^+$ goes to $\delta K^e$. We have plotted the sum of $U^-$, $\delta K^i$, and $\delta K^i$ in Fig.~\ref{fig:energy_p} (indicated in a green solid line), which shows that the total energy is nearly equal to $U^+$. A blue solid horizontal line in Fig.~\ref{fig:energy_p} indicates zero energy. The studies \cite{gonzalez_proton_2021,gonzalez_particle--cell_2023} with higher $\beta$ indicate that the ion heating is attributed to the phase front steepening of the waves and particles scattering at the steepened front. However, we do not find strong evidence of front steepening in either the forward or backward waves. {This is explained in Appendix ~\ref{sec:sup_mat} (Fig.~\ref{fig:front}(a))}. By contrast, a small degree of front steepening appears in the ion-acoustic waves {(Fig.~\ref{fig:front}(b))}, which is discussed later in this section. In addition to ion-acoustic wave steepening, ion heating may also result from ion Landau damping of ion-acoustic waves; this mechanism is also discussed later in this section. As inferred from Fig.~\ref{fig:energy_p}, electron heating is smaller than ion heating. It appears that electron heating arises from PDI, with the dominant contribution likely coming from acoustic wave steepening. However, we do not find {any} evidence of electron Landau damping of the acoustic waves, which may account for the relatively weak electron heating. \\

\begin{figure}
\includegraphics[width=8cm]{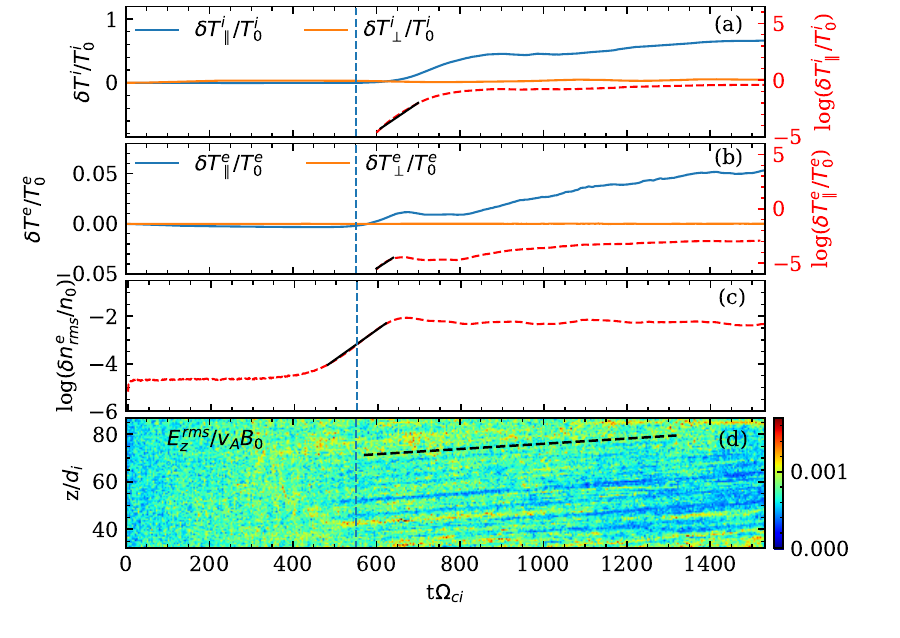}
\caption{\label{fig:tempral}Panels (a) and (b) show the ion and electron temperatures, while panels (c) and (d) show the root-mean-square (rms) electron plasma density fluctuation $\delta n_{\mathrm{rms}}^e/n_0$ and the parallel electric field $E_z^{\mathrm{rms}}/v_A B_0$, respectively. The black solid lines in panels (a)-(c) represent the growth rates of the corresponding quantities, while the black dashed line in panel (d) denotes the phase velocity of the ion-acoustic wave.}
\end{figure}

\indent To confirm that the heating of electrons and ions is primarily attributable to PDI, rather than to artifacts such as numerical noise or bulk plasma dynamics, we analyze the parallel and perpendicular temperatures of both the plasma species. {These temperatures are subtracted from their respective initial values}. The temporal evolution of the parallel and perpendicular ion temperature, $\delta T^i_{\parallel}/T^i_0$ and $\delta T^i_{\perp}/T^i_0$, is shown in Fig.~\ref{fig:tempral}(a) using blue and orange curves, respectively.  By contrast, the logarithm of $\delta T^i_{\parallel}/T^i_0$ is plotted as a red dashed line. We found that ion heating occurs predominantly in the parallel direction; $\delta T^i_\parallel/T^i_0$ increases when PDI starts growing, while $\delta T^i_\perp/T^i_0$ remains almost unchanged throughout. A solid black line is superimposed on the red dashed line to determine the growth rate of the ion energy ($\gamma_i$) in the parallel direction, yielding $\gamma_i \sim 0.024\Omega_{ci}$. A similar behavior is observed for the electron parallel and perpendicular temperatures, $\delta T^e_{\parallel}/T^e_0$ and $\delta T^e_{\perp}/T^e_0$, as shown in Fig.~\ref{fig:tempral}(b), where the logarithm of $\delta T^e_{\parallel}/T^e_0$ is also plotted as a red dashed line. The growth rate of the electron energy ($\gamma_e$), determined from the slope of the black solid line superimposed on $\log(\delta T^e_{\parallel}/T^e_0)$, is found to be $\gamma_e \sim 0.024\Omega_{ci}$. Furthermore, Fig.~\ref{fig:tempral}(c) shows the temporal evolution of the root-mean-square (rms) electron density fluctuation, $n^e_{\mathrm{rms}}/n_0$. An approximately exponential increase in $n^e_{\mathrm{rms}}/n_0$ over the interval $\delta t\Omega_{ci} = 450$-$650$ indicates the growth of PDI. The growth rate of PDI ($\gamma$) is determined to be $\gamma \sim 0.0125\Omega_{ci}$. Note that the ion density shows behavior similar to that of the electron density; hence, ion density fluctuations are not shown. We also present the spatio-temporal rms parallel electric field, $E^{\mathrm{rms}}_z / v_A B_0$, in Fig.~\ref{fig:tempral}(d). It is observed that finite fluctuations in $E^{\mathrm{rms}}z/v_A B_0$ begin to appear from $t \Omega_{ci} \sim 550$. This time, indicated by the vertical dashed line in all panels of Fig.~\ref{fig:tempral}, marks the onset of PDI as mentioned earlier in this section.  A black dashed line in \ref{fig:tempral}(d) is used to estimate the propagation speed of electric field fluctuations along the parallel direction, yielding a value of $\sim0.012v_A$, which is close to the ion-acoustic phase velocity estimated as $v_s=\omega_s/k_s \sim 0.025\Omega_{ci} / (2d_i^{-1}) \sim 0.0125v_A$. Here, $k_s$ denotes the ion-acoustic wave-vector obtained from the simulation. {Theoretically, ion acoustic speed can be calculated using $v_{sth}=\sqrt{(T_e+\Gamma_iT_i)/m_i }$ which yields $v_{sth}=0.0132 v_A$ for the given simulation parameters, Here, $\Gamma_i=3$ is used for ions}. The growth rates of the electron and ion energies are approximately twice the PDI growth rate, indicating that an interesting correlation between PDI and electron/ion heating. {This behavior can be understood from the fact that the PDI growth follows $\delta n(t) \sim e^{\gamma t}$. By contrast, the energy depends on the square of the fluctuation amplitude, $(\delta n(t))^2 \sim e^{2\gamma t}$. As a result, the particle energy growth rate is expected to be nearly twice the instability growth rate. This indicates that even a modest growth rate of instability can lead to very rapid energy transfer and particle heating.} In addition, finite fluctuations in the parallel electric field further support that electron and ion heating are strongly attributed to the PDI, predominantly along the ambient magnetic field. 

{Fig.~\ref{fig:tempral} shows a time lag between the onset of the PDI and the particle heating, indicating that significant particle heating requires sufficient development of the instability. As seen in Fig.~\ref{fig:tempral}(a), ion heating in the parallel direction occurs later than electron heating (Fig.~\ref{fig:tempral}(b)). This delay may be attributed to the lower electron mass, as the growth of the PDI (Fig.~\ref{fig:tempral}(c)) and the fluctuations in the parallel electric field (Fig.~\ref{fig:tempral}(d)) emerge well before observable particle heating. Consequently, electrons respond earlier than ions. It is also observed that, after the onset of the PDI ($t\Omega_{ci}\sim600$), both electrons and ions exhibit stronger heating, followed by a plateau in their energies, with a temporal offset attributable to the mass difference between the species. The initial heating phase is likely associated with the linear growth stage of the PDI, whereas the subsequent evolution is governed by multiple PDI cycles, as discussed earlier in this section (subsequent cycles in $Z^+$, $Z^-$, and $\delta n_e/n_0$). These later cycles are strongly nonlinear, making their detailed interpretation more complex. Beyond the plateau, both species continue to gain energy over time due to subsequent cycles of PDI. }

\begin{figure}
\includegraphics[width=8cm]{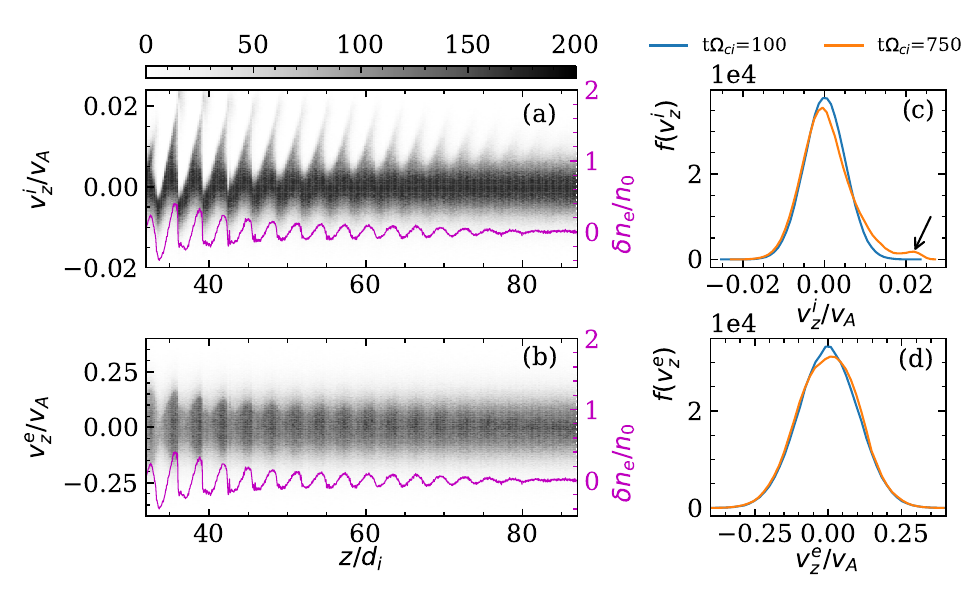}
\caption{\label{fig:particle} Panels (a) and (b) represent the ion and electron phase spaces { at time $t\Omega_{ci}=750$}, respectively, superimposed with electron plasma density fluctuations (magenta line), while panels (c) and (d) show the corresponding velocity distribution functions before and after PDI.}
\end{figure}

\indent Moreover, the ion and electron heating can be interpreted in terms of their phase-space structures in the parallel direction. The phase-space structures ($z$-$v_z$) of ions and electrons are shown in Fig.~\ref{fig:particle}(a) and (b), with the electron density perturbation ($\delta n_e/n_0$) overlaid as a magenta line. The alignment of density peaks with phase-space structures further strengthens the indication that PDI drives electron and ion heating. {This correlation suggests that the compressive fluctuations generated by PDI lead to localized enhancements in electric fields, which in turn facilitate wave-particle interactions such as trapping and phase mixing. These processes produce distortions in particle phase space, manifesting as phase-space structures, and enable efficient energy transfer from waves to particles, thereby resulting in plasma heating. } The corresponding ion and electron distribution functions, $f(v^i_z)$ and $f(v^e_z)$, are shown in Fig.~\ref{fig:particle}(c) and (d) for $t\Omega_{ci}= 100 $ and $t\Omega_{ci} = 750$, respectively. Fig.~\ref{fig:particle}(c) shows that a distinct bump (indicated by an arrow) in the ion distribution tail at $t\Omega_{ci} = 750$ indicates ion Landau damping of the ion-acoustic waves. We find that the observed bump appears near the ion-acoustic wave phase velocity, {which is $v_s\sim 0.0125 v_A$}. A nearly negligible modification of the electron distribution function, as shown in Fig.~\ref{fig:particle}(d) at $t\Omega_{ci}=750$, suggests that electron heating is not attributable to electron Landau damping of ion-acoustic waves, as the ion-acoustic phase velocity lies within the bulk region of the electron distribution. Instead, a small steepening of the acoustic waves is observed (Fig.~\ref{fig:particle}(a), magenta line), which can accelerate both electrons and ions, allowing them to gain energy from the acoustic waves. These mechanisms together suggest stronger ion heating and comparatively weaker electron heating. \\

\begin{figure}
\includegraphics[width=8cm]{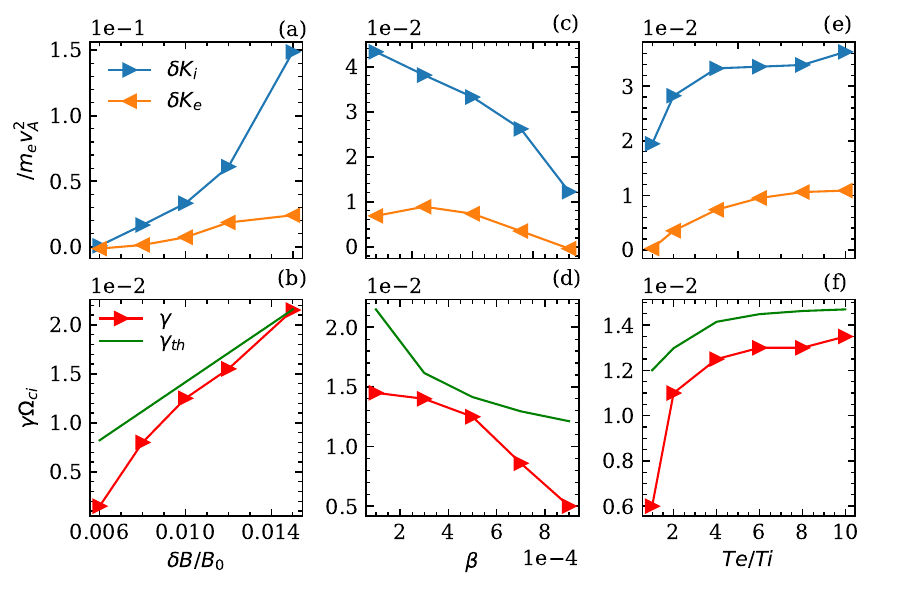}
\caption{\label{fig:growth_thermal_en} Panels (a), (c), and (e) show the behavior of ion and electron thermal energies as functions of $\delta B/B_0$, $\beta$, and $T_e/T_i$, respectively, while panels (b), (d), and (f) show the PDI growth rate ($\gamma$) calculated from simulations (red line) and $\gamma_{th}$ calculated from theory  (green line) as a function of $\delta B/B_0$, $\beta$, and $T_e/T_i$, respectively.}
\end{figure}

\indent To determine the scaling behavior and sensitivity to key plasma and wave parameters, we perform a parametric study of particle thermal energies and the PDI growth rate as functions of $\delta B/B_0$, $\beta$, and $T_e/T_i$. The increase in ion and electron thermal energies, $\delta K^i$ and $\delta K^e$ (shown in blue and orange, respectively), and the corresponding growth rate (shown in red and green for simulation and theory, $\gamma$ and $\gamma_{\mathrm{th}}$, respectively) are presented in Fig.~\ref{fig:growth_thermal_en}. Panels (a), (c), and (e) show $\delta K^i$ and $\delta K^e$, while panels (b), (d), and (f) show the growth rate, as functions of $\delta B/B_0$, $\beta$, and $T_e/T_i$, respectively. We have calculated the theoretical growth $\gamma_{\mathrm{th}}$ from the following expression \cite{li_measuring_2025},
\begin{align}
    \gamma_{\mathrm{th}}=\frac{1}{2}\left(-\Gamma_s+\sqrt{\Gamma_s^2+4\gamma_g^2} \right),
    \label{eq:pdi_g}
\end{align}
\noindent where $\Gamma_s$ and $\gamma_g$ denote the acoustic wave damping rate and ideal undamped growth rate, respectively. {The derivation of Eq.~\ref{eq:pdi_g} assumes the negligible damping of the backward AW}. Here, $\Gamma_s/\omega_s=1.1\times (T_e/T_i)^{-7/4} \exp(-(T_e/T_i)^{-2})$ and $\gamma_g/\omega_+=(1/2\sqrt2)\times (\delta B/B_0)\beta^{-1/4}$\cite{li_parametric_2022}. It is observed that $K^i$ and $K^e$ as well as $\gamma$, increase with $\delta B/B_0$ (Fig. \ref{fig:growth_thermal_en}(a) and (b)), decrease with $\beta$ (Fig.~\ref{fig:growth_thermal_en}(c) and (d)), and initially increase with $T_e/T_i$ and then saturate (Fig.~\ref{fig:growth_thermal_en}(e) and (f)). {The increase with $\delta B/B_0$ reflects the stronger nonlinear coupling associated with larger pump-wave amplitudes, which enhances the efficiency of the parametric decay instability (PDI) and the subsequent energy transfer to particles. The decrease with $\beta$ can be attributed to enhanced thermal effects at higher $\beta$, which weaken compressibility and increase damping (e.g., Landau damping), thereby suppressing PDI growth and reducing particle heating.} As per the given expressions of $\beta_e$ and $\beta_i$ at the end of Sec.\ref{sec:parameters}, $\beta_i$ decreases with $T_e/T_i$, while $\beta_e$ initially increases and then likely saturates. {The dependence on $T_e/T_i$ arises from the role of electron pressure in mediating the compressive daughter modes. When $T_e \sim T_i$, the growth rate is expected to be relatively low due to strong Landau damping of the ion-acoustic mode in this regime \cite{hasegawa_kinetic_1976}. The initial rise in the growth rate, followed by a plateau as seen in Fig.~\ref{fig:growth_thermal_en}(f), is therefore likely associated with the corresponding variation in the Landau damping rate of the acoustic mode, which decreases initially and then saturates at higher $T_e/T_i$.} We obtain a qualitatively good agreement between $\gamma$ and $\gamma_{\mathrm{th}}$.\\

\section{Summary and conclusions}\label{sec:summary}
In this work, we simulate AW PDI in one spatial dimension using a fully kinetic PIC approach with realistic absorbing boundary conditions. The parameters considered correspond to typical conditions in LAPD. The energy transfer from the pump wave to the daughter modes, as well as the partitioning between electrons and ions, is examined over a range of wave and plasma parameters.  Simulation results demonstrate that nearly 92\% of the pump energy is transferred to the counter-propagating AW, 5-6\% to ions, and approximately 1-2\% to electrons.  It is found that ion and electron heating occur predominantly in the parallel direction. Further results indicate that the parallel energy growth rates of ions and electrons are almost twice the PDI growth rate {which is attributed to the quadratic dependence of the energy growth on the fluctuation amplitude}. We observe weak electron heating and, consequently, an almost negligible feedback on the PDI. The present study will help in designing higher-dimensional simulations of AW PDI, where strong electron heating is expected to arise from electron Landau damping of AWs, making the role of kinetic electrons more prominent in the PDI dynamics. \\

\appendix
\renewcommand{\thefigure}{A\arabic{figure}}
\setcounter{figure}{0}
\section{Supplementary Materials} \label{sec:sup_mat}

\begin{figure}
\includegraphics[width=8cm]{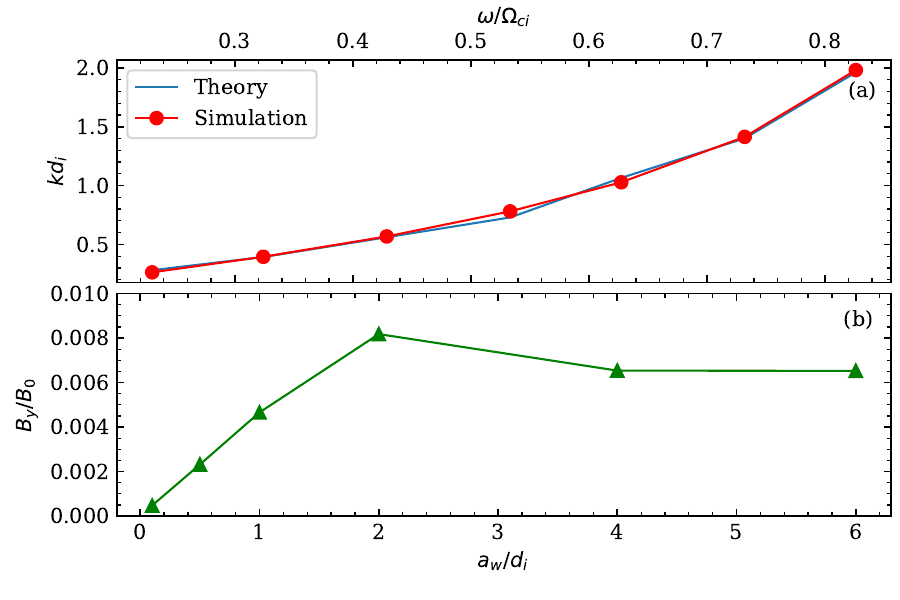}
\caption{\label{fig:disp} Panel (a) denotes the dispersion curve for a left-hand circularly Alfv\'en wave, while panel (b) represents the variation of pump wave amplitude ($B_y/B_0$) with wave injection width $a_w$.}
\end{figure}

We have verified the theoretical dispersion relation (Eq.~\ref{eq:disp}) for a left-hand circularly polarized AW against the dispersion obtained from the simulation using the set of parameters given in Sec.~\ref{sec:parameters}, as shown in Fig.~\ref{fig:disp}(a). In this figure, solid blue lines represent the theoretical curve, while red lines with markers show the simulation results. The two curves nearly overlap, demonstrating excellent agreement between theory and simulation. We have also confirmed that the right-hand circularly polarized AW follows the relation $\omega/k = v_A \sqrt{1 + \omega/\Omega_{ci}}$. Furthermore, Fig.~\ref{fig:disp}(b) shows the variation of the perpendicular magnetic field $B_y/B_0$ with the wave injection width $a_w$, indicated by green lines with markers. It is observed that $B_y/B_0$ initially increases with $a_w$ and subsequently saturates beyond $a_w = 2d_i$. Accordingly, we adopt $a_w = 2d_i$ for all simulations presented in this manuscript.\\

\begin{figure}
\includegraphics[width=8cm]{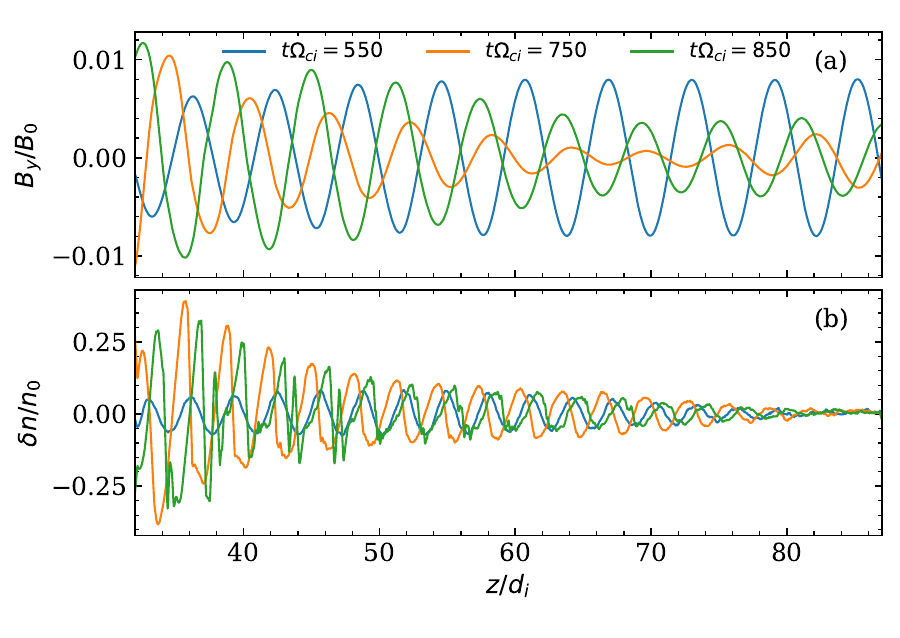}
\caption{\label{fig:front} Panels (a) and (b) show the perpendicular magnetic field, $B_y/B_0$, and plasma density fluctuations, $\delta n/n_0$, respectively, at times $t\Omega_{ci} = 550, 750,$ and $850$. Note that the above plot is shown only for the central region.}
\end{figure}

To investigate front steepening in AWs and ion-acoustic waves, the spatial variations of $B_y/B_0$ and $\delta n/n_0$ are shown at times $t\Omega_{ci} = 550, 750,$ and $850$ in the central region, as shown in Fig.~\ref{fig:front}(a) and (b), respectively.  It can be seen in Fig.~\ref{fig:front}(a) that the amplitude of $B_y/B_0$ decreases at $t\Omega_{ci} = 750$ and $850$ compared to $t\Omega_{ci} = 550$, which is attributed to PDI, while the overall shape remains nearly unchanged, indicating negligible front steepening in AWs. On the other hand, shapes of the $\delta n/n_0$ change with time (see Fig.~\ref{fig:front} (b)), indicating a clear front steepening of the ion-acoustic waves. 

\bibliographystyle{unsrt}
\bibliography{citation}

\end{document}